\documentclass[aps,
prl,
10pt,
twocolumn,
superscriptaddress,
amsmath,
amssymb,
]{revtex4-2}

\usepackage{graphicx} 
\usepackage{dcolumn}  
\usepackage{bm}       
\usepackage[colorlinks=true,citecolor=blue,filecolor=blue,linkcolor=blue,urlcolor=blue,pdftex]{hyperref}
\usepackage{color}
\usepackage{comment}
\usepackage{booktabs}
\usepackage{orcidlink} 

\begin{document}

\preprint{APS/123-QED}

\title{Search for Light Dark Matter in Low-Energy Ionization Signals from XENONnT}

\newcommand{\bologna}{\affiliation{Department of Physics and Astronomy, University of Bologna and INFN-Bologna, 40126 Bologna, Italy}}
\newcommand{\chicago}{\affiliation{Department of Physics, Enrico Fermi Institute \& Kavli Institute for Cosmological Physics, University of Chicago, Chicago, IL 60637, USA}}
\newcommand{\coimbra}{\affiliation{LIBPhys, Department of Physics, University of Coimbra, 3004-516 Coimbra, Portugal}}
\newcommand{\columbia}{\affiliation{Physics Department, Columbia University, New York, NY 10027, USA}}
\newcommand{\lngs}{\affiliation{INFN-Laboratori Nazionali del Gran Sasso and Gran Sasso Science Institute, 67100 L'Aquila, Italy}}
\newcommand{\mainz}{\affiliation{Institut f\"ur Physik \& Exzellenzcluster PRISMA$^{+}$, Johannes Gutenberg-Universit\"at Mainz, 55099 Mainz, Germany}}
\newcommand{\mpik}{\affiliation{Max-Planck-Institut f\"ur Kernphysik, 69117 Heidelberg, Germany}}
\newcommand{\munster}{\affiliation{Institut f\"ur Kernphysik, University of M\"unster, 48149 M\"unster, Germany}}
\newcommand{\nikhef}{\affiliation{Nikhef and the University of Amsterdam, Science Park, 1098XG Amsterdam, Netherlands}}
\newcommand{\nyuad}{\affiliation{New York University Abu Dhabi - Center for Astro, Particle and Planetary Physics, Abu Dhabi, United Arab Emirates}}
\newcommand{\purdue}{\affiliation{Department of Physics and Astronomy, Purdue University, West Lafayette, IN 47907, USA}}
\newcommand{\rice}{\affiliation{Department of Physics and Astronomy, Rice University, Houston, TX 77005, USA}}
\newcommand{\stockholm}{\affiliation{Oskar Klein Centre, Department of Physics, Stockholm University, AlbaNova, Stockholm SE-10691, Sweden}}
\newcommand{\subatech}{\affiliation{SUBATECH, IMT Atlantique, CNRS/IN2P3, Nantes Universit\'e, Nantes 44307, France}}
\newcommand{\torino}{\affiliation{INAF-Astrophysical Observatory of Torino, Department of Physics, University  of  Torino and  INFN-Torino,  10125  Torino,  Italy}}
\newcommand{\ucsd}{\affiliation{Department of Physics, University of California San Diego, La Jolla, CA 92093, USA}}
\newcommand{\wis}{\affiliation{Department of Particle Physics and Astrophysics, Weizmann Institute of Science, Rehovot 7610001, Israel}}
\newcommand{\zurich}{\affiliation{Physik-Institut, University of Z\"urich, 8057  Z\"urich, Switzerland}}
\newcommand{\paris}{\affiliation{LPNHE, Sorbonne Universit\'{e}, CNRS/IN2P3, 75005 Paris, France}}
\newcommand{\freiburg}{\affiliation{Physikalisches Institut, Universit\"at Freiburg, 79104 Freiburg, Germany}}
\newcommand{\napels}{\affiliation{Department of Physics ``Ettore Pancini'', University of Napoli and INFN-Napoli, 80126 Napoli, Italy}}
\newcommand{\nagoya}{\affiliation{Kobayashi-Maskawa Institute for the Origin of Particles and the Universe, and Institute for Space-Earth Environmental Research, Nagoya University, Furo-cho, Chikusa-ku, Nagoya, Aichi 464-8602, Japan}}
\newcommand{\laquila}{\affiliation{Department of Physics and Chemistry, University of L'Aquila, 67100 L'Aquila, Italy}}
\newcommand{\tokyo}{\affiliation{Kamioka Observatory, Institute for Cosmic Ray Research, and Kavli Institute for the Physics and Mathematics of the Universe (WPI), University of Tokyo, Higashi-Mozumi, Kamioka, Hida, Gifu 506-1205, Japan}}
\newcommand{\kobe}{\affiliation{Department of Physics, Kobe University, Kobe, Hyogo 657-8501, Japan}}
\newcommand{\kit}{\affiliation{Institute for Astroparticle Physics, Karlsruhe Institute of Technology, 76021 Karlsruhe, Germany}}
\newcommand{\tsinghua}{\affiliation{Department of Physics \& Center for High Energy Physics, Tsinghua University, Beijing 100084, P.R. China}}
\newcommand{\ferrara}{\affiliation{INFN-Ferrara and Dip. di Fisica e Scienze della Terra, Universit\`a di Ferrara, 44122 Ferrara, Italy}}
\newcommand{\groningen}{\affiliation{Nikhef and the University of Groningen, Van Swinderen Institute, 9747AG Groningen, Netherlands}}
\newcommand{\westlake}{\affiliation{Department of Physics, School of Science, Westlake University, Hangzhou 310030, P.R. China}}
\newcommand{\shenzhen}{\affiliation{School of Science and Engineering, The Chinese University of Hong Kong (Shenzhen), Shenzhen, Guangdong, 518172, P.R. China}}
\newcommand{\coimbrapoli}{\affiliation{Coimbra Polytechnic - ISEC, 3030-199 Coimbra, Portugal}}
\newcommand{\uniheidelberg}{\affiliation{Physikalisches Institut, Universit\"at Heidelberg, Heidelberg, Germany}}
\newcommand{\roma}{\affiliation{INFN-Roma Tre, 00146 Roma, Italy}}
\newcommand{\bucknell}{\affiliation{Department of Physics \& Astronomy, Bucknell University, Lewisburg, PA, USA}}
\author{E.~Aprile\,\orcidlink{0000-0001-6595-7098}}\columbia
\author{J.~Aalbers\,\orcidlink{0000-0003-0030-0030}}\groningen
\author{K.~Abe\,\orcidlink{0009-0000-9620-788X}}\tokyo
\author{S.~Ahmed Maouloud\,\orcidlink{0000-0002-0844-4576}}\paris
\author{L.~Althueser\,\orcidlink{0000-0002-5468-4298}}\munster
\author{B.~Andrieu\,\orcidlink{0009-0002-6485-4163}}\paris
\author{E.~Angelino\,\orcidlink{0000-0002-6695-4355}}\torino\lngs
\author{D.~Ant\'on~Martin\,\orcidlink{0000-0001-7725-5552}}\chicago
\author{F.~Arneodo\,\orcidlink{0000-0002-1061-0510}}\nyuad
\author{L.~Baudis\,\orcidlink{0000-0003-4710-1768}}\zurich
% ...
\collaboration{XENON Collaboration}\email[]{xenon@lngs.infn.it}\noaffiliation

\begin{abstract}
We report on a blinded search for dark matter with single- and few-electron signals in the first science run of XENONnT relying on a novel detector response framework that is  physics-model-dependent. We derive 90\% confidence upper limits for dark matter-electron interactions. Heavy and light mediator cases are considered for the standard halo model and dark matter up-scattered in the Sun. We set stringent new limits on dark matter-electron scattering via a heavy mediator with a mass within 10-20\,MeV/$c^2$ and electron absorption of axion-like particles and dark photons for $m_\chi$ below 0.186\,keV/$c^2$.
\end{abstract}

\maketitle

\textit{Introduction}.
Astronomical observations suggest the existence of a substantial amount of dark matter (DM) in our Universe ~\cite{Bertone:2004pz, Cirelli:2024ssz}. Many experiments aim to detect and clarify the nature of the hypothetical DM particle~\cite{MarrodanUndagoitia:2015veg, Roszkowski:2017nbc}. One of these is the XENONnT experiment~\cite{XENON:2024wpa}, which is optimized to search for DM-nucleon scattering for DM candidates in the GeV/$c^2$ to TeV/$c^2$ range~\cite{XENON:2023cxc}. In this Letter, we present the limits on light DM ($<$~GeV/$c^2$) candidates by analyzing the ionization signals observed within the detector. 

The XENONnT experiment consists of three nested detectors. The two outer detectors, both filled with deionized water, serve as a Cherenkov muon veto~\cite{XENON1T:2014eqx} and a neutron veto respectively~\cite{nt_nv}. A double-walled cryostat, containing a total of 8.5\,t of liquid xenon (LXe), holds the central dual-phase time projection chamber (TPC) wherein 5.9\,t of LXe is instrumented within a 1.49\,m tall and 1.33\,m diameter cylinder~\cite{XENON:2024wpa}. 

Particles scattering off xenon atoms result in both prompt scintillation photons, labeled signal 1 (S1) when detected, and ionization electrons. Ionization electrons are drifted to the top of the LXe by a 23\,V/cm drift field maintained by a cathode electrode at the bottom of the TPC and a gate electrode at the top. Electrons are then extracted into a gaseous xenon (GXe) phase by a 2.9\,kV/cm in-liquid extraction field maintained between the gate electrode and an anode electrode. Extracted electrons produce secondary scintillation light, labeled signal 2 (S2) when detected, proportional to the number of electrons. Scintillation photons are collected by photomultiplier tubes (PMTs) at the top and bottom of the detector. Pairing S1s and S2s allows for three-dimensional position reconstruction of the interaction location: the S1 and S2 arrival time difference provides the depth of the interaction ($Z$), while the S2 light pattern indicates the $XY$ position in the circular cross-section plane of the TPC.

DM searches requiring an S2 to be paired with an S1 for event reconstruction have set the most stringent limits on dark matter masses above $\sim$\,6\,GeV/$c^2$~\cite{XENON:2023cxc, LZ:2022lsv, PandaX:2024qfu, aprile2024searchlightdarkmatter}. For DM particles below this mass, elastic scattering interactions rarely produce detectable S1s, thus searches for such light DM candidates are conducted without requiring an S1~\cite{PhysRevLett.123.251801, PandaX-II:2021nsg, PandaX:2022xqx}. A large instrumental background of single- and few-electron S2s has been observed~\cite{Akerib:2020jud, XENON:2021qze} that persists $\mathcal{O}$(100)\,ms following particle interactions. We mitigate this background by developing stringent spatial and temporal vetos for the smallest S2s and limit our region of interest (ROI) to S2s with size subject to these veto requirements. 

We rely on two subsets of data from the initial science run of XENONnT (SR0)~\cite{XENON:2023cxc, XENON:2022ltv} to search for DM scattering off electrons, producing electronic recoils (ER). The first dataset (D1) with a live-time of 14.3 days is collected in November 2021 at the end of SR0 and is used to refine our data selections. The second dataset (D2) with 16.5 days of science data is collected in September and October of 2021, is used for the blinded search. To probe DM models producing ERs down to a single electron, we lower the recoil energy threshold of 1\,keV achieved in Ref.~\cite{XENON:2022ltv} to 13.7\,eV. Finally, we set limits on certain DM models using D2 and also report the results from D1.

\textit{Detector Response}.
To model a DM-electron interaction producing an ER with an initial energy $E_\mathrm{R}$, we use the best-fit detector response model for XENONnT~\cite{xenoncollaboration2024xenonntwimpsearchsignal}. We assume conservatively that the initial number of quanta (photons or electrons)  produced by the interaction are given by integer $N_q = E_\mathrm{R}/W$, where $W=13.7$\,eV~\cite{Dahl:2009nta} is the energy required to create a single quantum. The number of ionization electrons is modeled by a random variable selected from a binomial process with $N_q$ trials and success probability of 0.88. While drifting to the gate electrons are lost due to attachment to electronegative impurities within the LXe. To account for this loss we apply an average attenuation factor of 0.07\,(0.11) during the D2 (D1) period, as derived from the measured electron loss throughout SR0~\cite{xenoncollaboration2024xenonntanalysissignalreconstruction}. The number of electrons extracted into the GXe is determined by the measured extraction efficiency of 53\%~\cite{XENON:2022ltv} and is referred to in this work as the true number of electrons $N_{\mathrm{e,true}}$. The secondary scintillation light produced by an electron in the GXe, as a function of the number of photoelectrons (PE) observed by the PMTs, is modeled as a Gaussian-Poisson mixture distribution with a Gaussian mean of $\mu=\mu_{1e} N_\mathrm{e, true}$ and a standard deviation of $\sigma=\sigma_{1e}\sqrt{N_\mathrm{e, true}}$~\cite{Ionization:12344}, where $
\mu_{1e}= 31.15$\,PE is our measured single electron gain with standard deviation $\sigma_{1e} = 6.62$\,PE. 

For S2s in our ROI, the exact value of $N_{\mathrm{e,true}}$ has important impacts on the expected temporal extent (width) of the recorded S2 signal. This is because the arrival times of individual electrons in few-electron S2s are dispersed as a result of the diffusion of electrons while drifting (up to 2.2 ms from the bottom of the TPC). Single-electron S2s by definition do not experience diffusion. Therefore, to more accurately estimate the number of electrons in an S2, we split the S2s into subcomponents based on local minima in the combined digitized response from all PMTs, and quantify the integer number of electrons in each subcomponent by its size in PE. The sum of electron contributions in all subcomponents for each S2 is referred to as $N_{\mathrm{\mathrm{e,rec}}}$ in this work.

\begin{figure}[!t]
    \centering
    \includegraphics[width=\columnwidth]{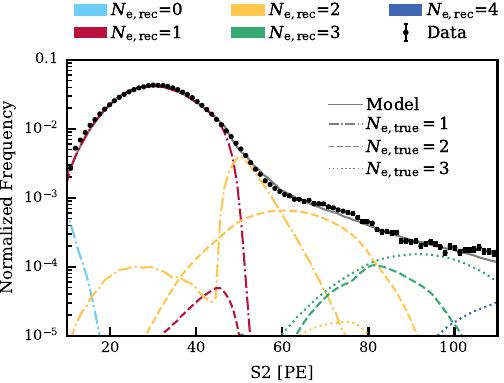}
    \caption{The observed single- and few-electron S2 (black data) distribution in SR0 D1. The data is compared to simulated $N_{\mathrm{e,true}}$ signals from 1, 2 and 3 electrons, represented as s dash-dotted, dashed and dotted curves respectively. For each $N_{\mathrm{e,true}}$, reconstruction artifacts result in a spread in observed $N_{\mathrm{e,rec}}$ values, which are then projected onto the S2 parameter space. S2s with $N_\mathrm{e,rec}=0$ originate from under-fluctuations in size, resulting in no subcomponent being large enough to be considered a reconstructed electron.}
    \label{fig:ne_true_to_rec_2}
\end{figure}

Reconstruction artifacts can produce a mismatch between $N_{\mathrm{e,rec}}$ and $N_{\mathrm{e,true}}$. The value of $N_{\mathrm{e,rec}}$ obtained from an observed S2 with $N_{\mathrm{e,true}}>1$ depends on fluctuations of the observed   S2 size (in PE) and the depth-dependent diffusion of its electrons. Without S1s, we cannot reconstruct the depth of events; therefore, we simulate S2s of different $N_{\mathrm{e,true}}$ throughout the drift region, and encode the probability of obtaining ($N_{\mathrm{e,rec}}$, S2) given $N_{\mathrm{e,true}}$~\cite{Ionization:12344} in a tensor $\mathcal{T}(N_{\mathrm{e,true}}, ~N_{\mathrm{e,rec}}, ~\text{S2})$. In Fig.~\ref{fig:ne_true_to_rec_2}, the effect of this decomposition is shown, where the observed size distribution of single- and few-electron S2s (black data) is compared to simulated signals ranging from $N_{\mathrm{e,true}}=1$ to 3. We also show the components in $N_{\mathrm{e,rec}}$ space that contribute to the observed spectrum.

\textit{Data Selection}.
We divide our data into windows of exposure defined by the time between two large ($N_{\mathrm{e,rec}}>$10) S2s, referred to as primary S2s (pS2s), within which we search for single- and few-electron S2s. This is similar to the procedure reported in Ref.~\cite{XENON:2021qze}. We consider DM interactions via ER signals of five true electrons or fewer, which would produce S2s within an analysis ROI of $\mathrm{S2}\in[10, 174]$\,PE. The applied data quality criteria are divided into two classes: peak quality selection criteria and exposure selections, where the latter select volumes and time periods in the detector with minimal backgrounds. 

\paragraph{Peak Quality Selection}
S2s are generated in the GXe region near the detector's top, where on average 75\% of photons are observed by the top PMT array. We calculate the likelihood of observing a given fraction of reconstructed PEs in the top PMT array for a given S2 size and set a threshold (``top fraction'') that retains 99\% of simulated S2s. The light patterns of S2s on the top PMT array are compared to patterns observed in $^{83\mathrm{m}}$Kr calibration events, with a likelihood score subsequently assigned~\cite{xenoncollaboration2024xenonntanalysissignalreconstruction}. A PE-dependent threshold on this likelihood score  (``hit pattern''), set to retain 98\% of  events from $^{220}$Rn calibration in our ROI, is then applied on the science search data.

\begin{figure}[!t]
    \centering
    \includegraphics[width=\columnwidth]{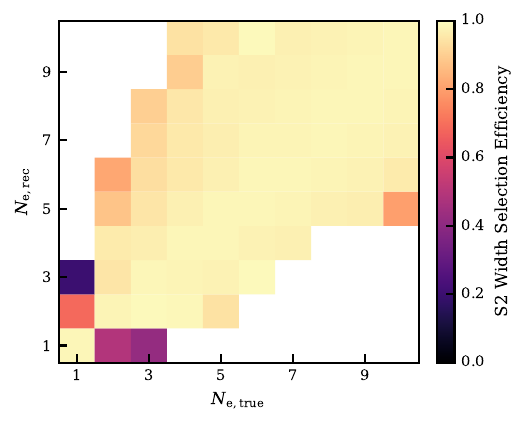}
    \caption{S2 width selection efficiency as a function of $N_{\mathrm{e,true}}$ and $N_{\mathrm{e,rec}}$. For large $N_{\mathrm{e,true}}$ efficiencies approach 98\%. Single-electron S2s have tight, drift-time independent selection thresholds. For multiple electron S2s that experience diffusion, selection thresholds are much larger and broader than for single-electron S2s. Therefore a significant drop in acceptance results when either $N_{\mathrm{e,true}}>1$ and $N_{\mathrm{e,rec}}=1$, or $N_{\mathrm{e,rec}}>1$ and $N_{\mathrm{e,true}}=1$.}
    \label{fig:ne_true_to_rec_1}
\end{figure}

To reject contemporaneous S2s or misidentified S1s, we simulate S2s over the full drift length and develop 98\% selection criteria on S2 width in $N_{\mathrm{e,true}}$ space (``width''). Due to reconstruction artifacts~\cite{Ionization:12344}, a constant-efficiency selection on width in $N_{\mathrm{e,true}}$ space results in localized drops in efficiency after converting to width in $N_{\mathrm{e,rec}}$ space. Fig.~\ref{fig:ne_true_to_rec_1} shows the selection efficiency as a function of $N_{\mathrm{e,true}}$ and $N_{\mathrm{e,rec}}$. The S2 widths of single-electron events are drift-time independent and have narrow selection bounds, whereas multiple-electron S2s diffuse while drifting, requiring wider selection bounds. This difference produces drops in selection efficiency for events with $N_{\mathrm{e,true}}>1$ and $N_{\mathrm{e,rec}}=1$, or $N_{\mathrm{e,rec}}>1$ and $N_{\mathrm{e,true}}=1$. 

Anomalous signals (e.g. from afterpulses) in the PMTs can bias $N_{\mathrm{e,rec}}$ upwards. In S2s with one true electron this mismatch between $N_{\mathrm{e,rec}}$ and $N_{\mathrm{e,true}}$ affects the signal acceptance. We apply a PE-dependent minimum time gap criterion between the first and second reconstructed electron subcomponents of an S2 (``after pulse''), as the timescale of anomalous light emission is $\mathcal{O}(10)$ times smaller than electron diffusion. The resulting detection efficiency of 96\% (99\%) for $N_{\mathrm{e,rec}}=2$ (3) presents negligible loss for larger signals. For S2s with $N_{\mathrm{e,true}}=1$ we account for a 1\% probability of the signal being observed with anomalous light emission, and thus removed by this selection. Finally, we reject peaks for which the number of contributing PMTs is larger than the S2 size in PE, as this is from misreconstruction due to electronic noise.

\begin{figure}[!t]
    \centering
    \includegraphics[width=\columnwidth]{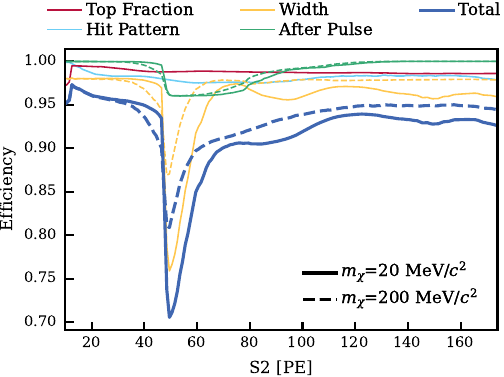}
    \caption{Selection efficiencies versus S2 size in PE. The PE dependent criteria, consisting of S2 hit pattern (cyan) and S2 area fraction top (red), are model-independent. Selections to remove S2s with after pulses ($N_{\mathrm{e,rec}}$-dependent, green) and based on the S2 widths ($N_{\mathrm{e,true}}$-dependent, yellow) are model-dependent. Cut efficiencies for DM-electron scattering via heavy mediator are shown for two DM masses (solid for $m_\chi = 20\,\mathrm{MeV}/c^2$, dashed for $m_\chi = 200\,\mathrm{MeV}/c^2$). The expected efficiency loss at 48\,PE originates from the known efficiency loss for the S2 width selection shown in Fig.~\ref{fig:ne_true_to_rec_1}.}
    \label{fig:cut_efficiencies}
\end{figure}

The selection efficiencies in their respective parameter spaces are accounted for in $\mathcal{T}$, and are shown in Fig.~\ref{fig:cut_efficiencies}, where it is noticeable that the signal efficiency is model-dependent when projected onto the ROI in S2 space. This effect is driven by the changes in the expected relative rate of single- and few-electron S2s across various DM models, and therefore the contribution of the selection efficiency loss when $N_{\mathrm{e,true}}$ and $N_{\mathrm{e,rec}}$ are mismatched also varies.

\paragraph{Exposure Selections} 
Significant background rates of single- and few-electron S2s in our ROI likely originate from photoionization of electronegative impurities, and delayed electron emission after higher-energy interactions~\cite{XENON:2021qze}. Thus, we restrict our search region to time periods and volumes of the detector that minimize the rate of these backgrounds. These selections were optimized using D1. We select events within a maximum radius of 39.5 cm from the center of the detector to reduce backgrounds from low-energy nuclear decays at the walls~\cite{xenoncollaboration2024xenonntanalysissignalreconstruction}. XENONnT's electrodes are comprised of parallel wires. The wire grid of both the gate and the anode is supported by two wires aligned perpendicular to the grid direction and passing horizontally across the TPC. An increased rate of single- and few-electron S2 signals is observed from the location of these support wires. We exclude a 15\,cm wide band around each perpendicular wire location. Intermittently throughout SR0, localized elevated emission of single-electron signals was observed in close proximity to the perpendicular wires, referred to as a ``hot spot''. We apply a fixed radius veto of 15\,cm centered at this position. 

We observe that the emission of single and few-electron signals are closely associated in position to a previous pS2. We therefore apply an exclusion radius around the position of any pS2 for the duration of the corresponding exposure window. This exclusion radius ranges from 20\,cm for S2s at our energy threshold of 10\,PE, to approximately 5\,cm at 174\,PE due to increased position reconstruction accuracy with larger S2s~\cite{xenoncollaboration2024xenonntanalysissignalreconstruction}. 

The position correlation between small S2s and their preceding pS2s are unclear when a cluster of pS2s occurs. Therefore, we remove the exposure windows following all pS2s in the cluster. We define a ``cluster'' by taking into account both the relative size of each pS2 and their proximity in time. Delayed electron emission from one pS2 is often observed after subsequent pS2s, even if the two interactions are not in close time proximity. This results in ``uncorrelated'' delayed electron emission within an exposure window. Therefore, we veto the first few hundred milliseconds following any pS2 in the detector, with the duration of the veto set by optimizing the signal to noise ratio. The veto duration decreases with the $N_{\mathrm{e,rec}}$ of the S2, and increases with the size of the pS2. Finally, as delayed electron emission is assumed to originate from particle interactions within the LXe~\cite{XENON:2021qze}, we require the interaction that produced the pS2 to have occurred within the LXe and be well-reconstructed, otherwise the entire subsequent exposure window is vetoed.

\begin{figure}[!t]
    \centering
    \includegraphics[width=\columnwidth]{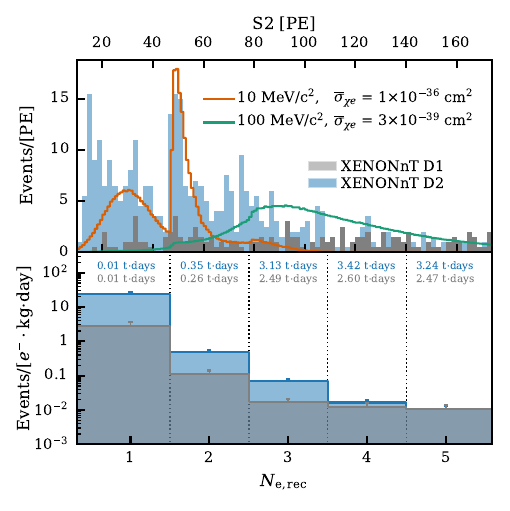}
    \caption{Top: Events that pass all selection criteria in both the blind search (gray) and unblind analysis (blue) relying on the D2 and D1 respectively. The expected detector response for DM-e scattering assuming a heavy mediator with masses of $m_\chi = 10\,\mathrm{MeV}/c^2$ (green) and $m_\chi = 100\,\mathrm{MeV}/c^2$ (orange) are also illustrated. Bottom: The measured rate of single- and few-electron S2s in D1 and D2 and corresponding exposures.}
    \label{fig:event_rates}
\end{figure}

\paragraph{Science Data}
The exposure selections result in the live-time of the data used in the DM search being smaller than the raw live-time of 16.4 days for D2. Additionally, the allowed volume changes for the exposure window of each pS2 and theoretically observed $\mathrm{S2}_{\mathrm{rec}}$. We correct for distortions of the drift field, which result in smaller observed radial positions at larger depths within the TPC, by multiplying the geometric volume with a correction factor of 1.19 obtained from inverting the field-distortion correction as introduced in Ref.~\cite{xenoncollaboration2024xenonntanalysissignalreconstruction}. The live-time after each pS2 is determined by the $N_{\mathrm{e,rec}}$ dependent time veto.  Exposure, summarized in Fig.~\ref{fig:event_rates}, increases with increasing $N_{\mathrm{e,rec}}$, primarily due to less stringent time veto selections following pS2s, and subdominantly due to smaller exclusion radii as S2 size increases.

The unblinded events observed in D2 are shown in Fig.~\ref{fig:event_rates}. The expected signal from DM-electron scattering, based on our detector response model and assuming a heavy mediator, is shown for DM masses of $m_\chi = 10~\mathrm{MeV}/c^2$ and $m_\chi = 100~\mathrm{MeV}/c^2$. For DM masses expected to produce predominantly S2s with $N_\mathrm{e,true}=1$, we expect a peak at the transition from $N_\mathrm{e,rec}=1$ to $N_\mathrm{e,rec}=2$ due to the $\mathcal{O}(30)$ times larger exposure in the latter bin, thereby amplifying the contribution of S2s with $N_\mathrm{e,true}=1$ and $N_\mathrm{e,rec}=2$. 

After unblinding, the observed single- and few-electron rate in D2 is higher than the expectation from D1 (training data). XENONnT was operated in a different mode, expected to contain more impurities, during the D1 period. The getter used to purify the GXe volume, which includes the GXe from above the detector as well as from two umbilical pipes which connect the cryostat to support infrastructure outside the muon veto water tank, was bypassed and GXe was returned directly to the cooling tower where it is reintroduced into the LXe volume of the TPC. Contrary to expectations, an exponential decrease in the rate of delayed electron emission is observed after bypassing the GXe getter. In addition, an excess of events with S2$<$20\,PE are observed in D2 near the ``hot spot'' exclusion region. These are single-electron S2s with under-fluctuations in size, from the known ``hot spot'' in SR0, and consequently suffer from position reconstruction accuracy artifacts larger than our exclusion veto.

Therefore, we performed a second search (after unblinding), reversing the role of the two datasets. Quality selections are independent of background rate and remain unchanged. Two of the exposure selections were retrained on the unblinded D2: the maximum radius and the veto window selections. The fixed D2 ``hot spot'' exclusion region was also changed to a PE-dependent exclusion veto similar to the already applied position correlation exclusion veto. The exposure, the final events and event rates shown in Fig.~\ref{fig:event_rates} for D1 are the result of this re-optimisation. 

\begin{figure*}[!tp]
\centering
\includegraphics[width=\textwidth]{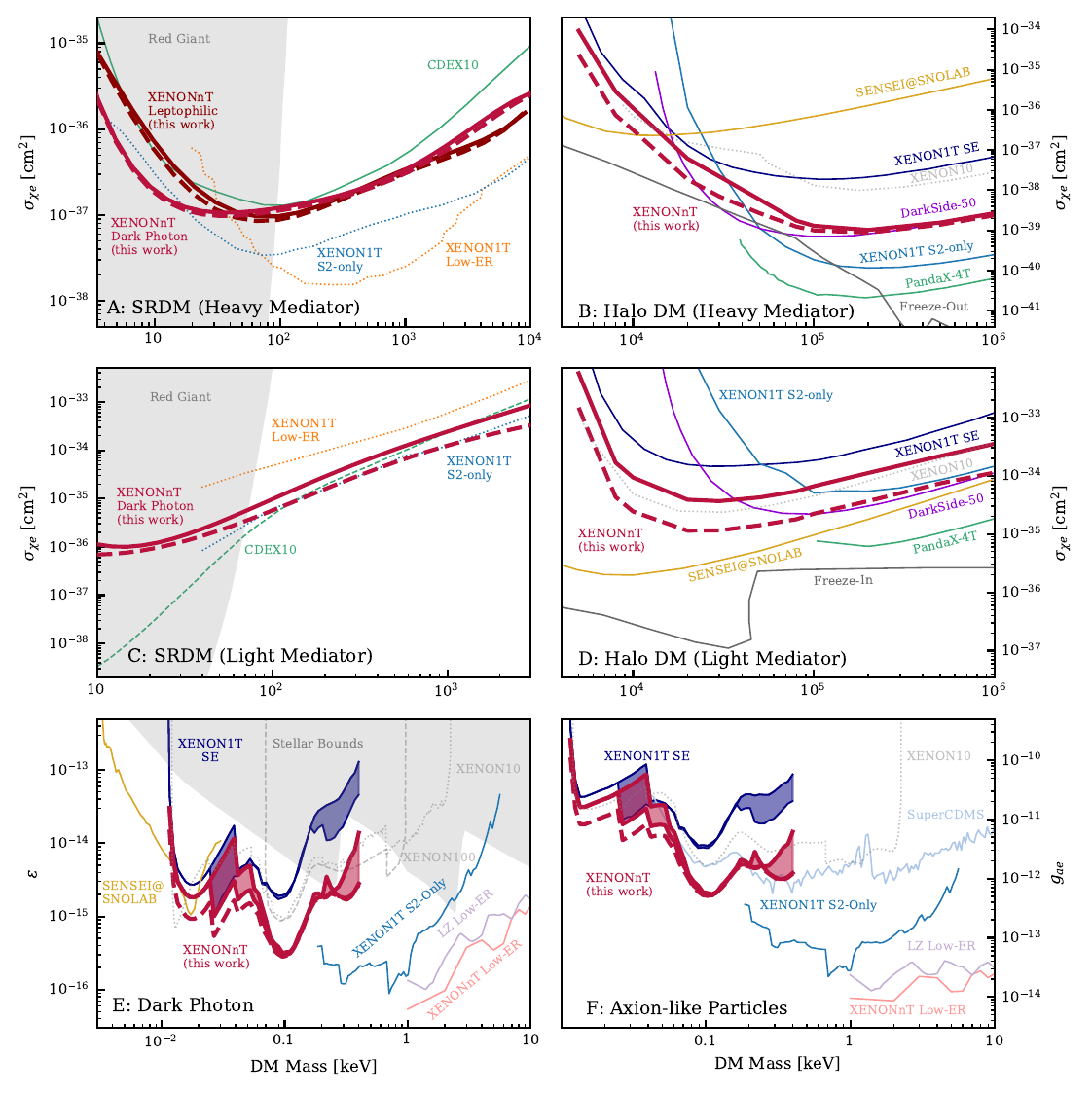}
\caption{Solid (dashed) red lines show the 90\% confidence level upper limits on the various physics model parameters computed using D2 (D1). Results calculated indirectly from previous experimental results are shown in dotted lines. Panel A (C) show the upper limits on DM-electron scattering cross section by considering solar-reflected DM under the heavy (light) mediator scenario, while panels B (D) show the limits by considering halo DM under the heavy (light) mediator scenario. Panel E shows the 90\% confidence level upper limits on the kinetic mixing parameter computed by considering Dark Photon models, while panel F shows the 90\% confidence level upper limits on the axio-electron coupling by considering Axion-like Particle models. We also show results from XENON10~\cite{PhysRevLett.107.051301}, XENON1T~\cite{XENON:2021qze,PhysRevLett.123.251801}, XENONnT Low-ER analysis~\cite{XENON:2022ltv}, LZ~\cite{PhysRevD.108.072006}, PandaX~\cite{PandaX-II:2021nsg,PandaX:2022xqx}, DarkSide-50~\cite{PhysRevLett.130.101002}, SuperCDMS~\cite{PhysRevD.101.052008}, SENSEI~\cite{SENSEI:2023zdf}, CDEX10~\cite{PhysRevLett.132.171001} as well as theoretical recasts of limits in dotted lines~\cite{Bloch_2017,srdm_heavy,PhysRevD.104.103026}, and the values of the model parameters required to obtain the correct relic abundance from freeze-out and freeze-in mechanisms in solid gray lines~\cite{essig2016direct}. The gray shaded areas in panels A, C, and E represent the parameter space excluded by stellar cooling constraints from red giants~\cite{srdm_rg_Chang_2021} and other stars~\cite{AN2015331}.}
\label{fig:limits}
\end{figure*}
\textit{Dark Matter Models}.
\label{section:DM_models}
Various DM models expected to result in ERs in our ROI are investigated. Light DM scattering off xenon orbital electrons with the physical model (form factor and structure function) described in Ref.~\cite{PhysRevLett.123.251801, XENON:2021qze}, is considered for the cases of heavy and light mediators. The ionization form factors are derived from Ref.~\cite{yu}. Two classes of DM particles are examined in this work: halo DM with a velocity distribution from the Standard Halo Model ~\cite{PhysRevLett.123.251801, XENON:2021qze, ddwhite_Baxter_2021}, and Solar-Reflected DM (SRDM) which are halo DM particles upscattered by the Sun~\cite{srdm_heavy, srdm_light}. 

We also consider bosonic DM candidates such as Dark Photon (DP) and Axion-like Particles (ALPs). DP can interact with electrons through a kinetic mixing (denoted by $\epsilon$) with the SM photon~\cite{HOLDOM1986196, GALISON1984279}, while ALPs are absorbed by bound electrons through the axioelectric effect~\cite{PhysRevD.35.2752, Pospelov_2008} (with axion-electron coupling strength $g_\mathrm{ae}$). Both models result in mono-energetic depositions corresponding to particle masses.

\textit{Results}.
\label{section:inference_results}
We observe 634 (189) events in the analysis ROI in D2 (D1). We compute the 90\% confidence level upper limits on the physics parameters of the various DM models~\cite{Bloch_2017, PhysRevD.96.043017} using the optimal interval method with the $p_\text{Max}$ test statistic~\cite{Yellin_2002} on the S2 size of the observed events. The trial factor incurred from testing all intervals within the dataset is accounted for via Monte Carlo simulations. The results are shown in Fig.~\ref{fig:limits}. The upper limits in bold solid (dashed) lines are computed using D2 (D1).

Panels A-D in Fig.~\ref{fig:limits} show the 90\% confidence level upper limits on the DM-electron cross section, $\sigma_{\chi e}$,  as a function of the DM mass, obtained in this work by considering the DM-electron scattering of both the halo and solar reflected DM. Existing upper limits in the literature include those from XENON1T~\cite{XENON:2021qze,PhysRevLett.123.251801}, other direct DM detection experiments~\cite{PhysRevLett.132.171001, SENSEI:2023zdf, PandaX-II:2021nsg, PandaX:2022xqx, PhysRevLett.130.101002}, and recasts using XENON1T data~\cite{srdm_heavy,PhysRevD.104.103026}. The solid gray lines in panels B and D indicate the values of the DM-electron cross section required to obtain the correct relic abundance from freeze-out and freeze-in production, respectively~\cite{essig2016direct}.

For SRDM, we are sensitive to DM masses down to 2 keV and 10 keV under the heavy and light mediator scenarios, respectively. The upper limits obtained in this work are stronger than the ones reported directly by previous experiments but less stringent than the recasts using XENON1T ionization-only data~\cite{srdm_heavy} and XENON1T Low-ER data~\cite{PhysRevD.104.103026} in the heavy mediator scenario as shown in panel A due to the choice of analysis ROI~\cite{Tan1853295}. For DM $\lesssim 100$ keV, constraints derived from the cooling of red giants (RG) have already ruled out, for both the heavy mediator and light mediator scenarios~\cite{srdm_rg_Chang_2021}, the region which our analysis is most suitable for.

Panel E shows the 90\% confidence level upper limits on the kinetic mixing parameter, $\epsilon$, by considering the DP model. Systematic uncertainty arises due to DP absorption in different electron shells. We calculate an upper limit, assuming absorption by either the outermost shell or the lowest energetically accessible shell, and shade the region between these extrema. We also present the constraints from XENON10~\cite{PhysRevLett.107.051301}, XENON100~\cite{Bloch_2017}, XENON1T~\cite{XENON:2021qze,PhysRevLett.123.251801}, XENONnT Low-ER analysis~\cite{XENON:2022ltv}, other direct DM experiments~\cite{PhysRevD.108.072006,SENSEI:2023zdf}, and stellar cooling constraints from the Sun, horizontal branch (HB) stars and RGs in solid gray lines~\cite{AN2015331}. 

The 90\% confidence level upper limits on the axion-electron coupling strength, $g_\mathrm{ae}$, computed by considering the ALP model is shown in panel F. The ALP parameter space is also constrained by XENON10~\cite{PhysRevLett.107.051301}, XENON1T~\cite{XENON:2021qze,PhysRevLett.123.251801}, XENONnT~\cite{XENON:2022ltv}, and other direct DM detection experiments~\cite{PhysRevD.108.072006,PhysRevD.101.052008}. The red shaded area represents the systematic uncertainty due to the unknown relative absorption rates by bound electrons in different xenon electron shells.

Our constraints on the DM-electron cross section are weakened by the elevated single-electron background rates in the BKG dataset. This only affects masses where the expected signal is concentrated in the 1-2 electron populations. DP and ALP models with masses heavier than 0.1 keV produce more electrons and are hence less affected by the increased rates of single-electron background events in the BKG dataset. This is can be seen from panels E-F in Fig.~\ref{fig:limits} from the convergence of the upper limits computed from D1 and D2 for larger DP or ALP masses.

We have reported the result of a blind search for light DM using single- and few-electron ionization signals in XENONnT with a novel detector response model. We exclude new parameter space for absorption of bosonic DM and DM electron scattering. Future work can improve on these results through the development of a predictive background model for delayed electron emission which will allow leveraging the full XENONnT exposure to probe new parameter space.

\textit{Acknowledgments}.
We gratefully acknowledge support from the National Science Foundation, Swiss National Science Foundation, German Ministry for Education and Research, Max Planck Gesellschaft, Deutsche Forschungsgemeinschaft, Helmholtz Association, Dutch Research Council (NWO), Fundacao para a Ciencia e Tecnologia, Weizmann Institute of Science, Binational Science Foundation, Région des Pays de la Loire, Knut and Alice Wallenberg Foundation, Kavli Foundation, JSPS Kakenhi, JST FOREST Program, and ERAN in Japan, Tsinghua University Initiative Scientific Research Program, DIM-ACAV+ Région Ile-de-France, and Istituto Nazionale di Fisica Nucleare. This project has received funding/support from the European Union’s Horizon 2020 research and innovation program under the Marie Skłodowska-Curie grant agreement No 860881-HIDDeN.

We gratefully acknowledge support for providing computing and data-processing resources of the Open Science Pool and the European Grid Initiative, at the following computing centers: the CNRS/IN2P3 (Lyon - France), the Dutch national e-infrastructure with the support of SURF Cooperative, the Nikhef Data-Processing Facility (Amsterdam - Netherlands), the INFN-CNAF (Bologna - Italy), the San Diego Supercomputer Center (San Diego - USA) and the Enrico Fermi Institute (Chicago - USA). We acknowledge the support of the Research Computing Center (RCC) at The University of Chicago for providing computing resources for data analysis.

We thank the INFN Laboratori Nazionali del Gran Sasso for hosting and supporting the XENON project.

\bibliography{biblio}

\end{document}